# Liquid photonic-molecule microlasers for ultrasensitive biosensing

Yan Wang,[1,2,3] Yu-Hao Hu,[1,2,3] Jin-Lei Wu,[4] Rui Duan,[5] Ya-Feng Jiao[1,2,3], Hai-Yan Wang,[1,2,3] Li-Ying Jiang,[1,2,3] Le-Man Kuang,[2,6,7] Han-Dong Sun,[8] and Hui Jing[2,6,7,9*]

[1]*School of Electronics and Information, Zhengzhou University of Light Industry, Zhengzhou 450001, China*

[2]*Academy for Quantum Science and Technology, Zhengzhou University of Light Industry, Zhengzhou 450001, China*

[3]*Henan Key Laboratory of Information Functional Materials and Sensing Technology, Zhengzhou University of Light Industry, Zhengzhou 450001, China*

[4]*Quantum Information Institute, School of Physics, Zhengzhou University, Zhengzhou 450001, China*

[5]*College of Physics, Sichuan University, Chengdu 610000, China*

[6]*Key Laboratory of Low-Dimensional Quantum Structures and Quantum Control of Ministry of Education, Hunan Normal University, Changsha 410081, China*

[7]*Department of Physics and Synergetic Innovation Center for Quantum Effects and Applications, Hunan Normal University, Changsha 410081, China*

[8]*Institute of Applied Physics and Materials Engineering, University of Macau, Macao SAR 999078, China*

[9]*Institute for Quantum Science and Technology, College of Science, National University of Defense Technology, Changsha 410073, China*

[*]Corresponding author. Email: jinghui@hunnu.edu.cn

## Abstract

Droplet microlasers, as promising tools for biophotonics and biomedical sciences, have witnessed rapid advances due to their flexible reconfigurability, high sensitivity to stimuli, and label-free biosensing ability. However, designing these biosensors with simultaneously critical properties of low lasing threshold, high spectral purity, and ultimate sensitivity remains challenging. Here, we propose a versatile strategy to build liquid photonic molecules (LPMs) that combine all these features in a single device. We find that through tailoring the spectral Vernier overlap in size-mismatched droplets, this device enables single-mode lasing with a low threshold of ~610 nJ mm$^{-2}$. The LPM lasers are engineered for dynamic tunability using a molecular isomerization strategy, which induces spectral mode hopping and thus yields a nearly ten-fold enhancement in spectral sensitivity over single droplets. Moreover, by leveraging the self-referenced intensity response of the LPM lasing modes, we demonstrate a three-orders-of-magnitude enhancement in biomolecular sensing, with a detection limit of 30 aM and a dynamic range spanning nine orders of magnitude. Our work offers exciting prospects for bio-integrated liquid sensors in diverse applications.

## 1. Introduction

Microlasers are key cornerstones of integrated photonics, optical communication, and modern life sciences[1–4]. Whispering-gallery-mode (WGM) microlasers, featuring spectrally narrow emission lines due to an extremely high quality ($Q$) factor and cavity resonance modulation[5], have emerged as an appealing platform for optical label-free detection and precise measurements[6–15]. With unique features of high flexibility, tunability, and reconfigurability, droplet WGM microlasers have recently garnered renewed interest since their first debut in the 1980s[16]. Although droplet microlasers are far more susceptible to environmental perturbations than their solid counterparts, they are, conversely, more sensitive to analytes[17,18], making them exceedingly viable as highly sensitive biosensors. Moreover, droplets possess a natural advantage over solids in their encapsulation capability, which enables high-throughput bioreaction analysis by integrating with microfluidic technology[19]. These merits have driven significant progress in droplet microlasers for biological applications, such as cell tagging and tracking, imaging, intracellular and deep tissue measurements[20–25]. However, most processes rely on the readout of characteristic behaviors (e.g., resonance shift, laser threshold and intensity) in the multimode laser spectra of droplets that possess low spectral purity and beam quality. The sensitivity is thus largely limited due to the dispersion of spectral responses to external stimuli across multiple modes. In addition, a lower lasing threshold is continually pursued for bioapplications to minimize thermal damage and phototoxicity[21,26], which is crucial for long-term live monitoring.

Mode selection through coupling multiple resonators has long been demonstrated as an effective means for constructing single-mode microlasers[27–31]. In the sense of physics, a strongly coupled photonic system comprising two (or more) size-mismatched resonators, known as photonic molecule (PM)[32], create super-modes extending over the coupled resonators. Compared to the original modes in individual resonators, the super-modes from collective multi-resonator resonances offer enhanced lasing and sensing capabilities[33–35]. Despite great advancements, the construction of PMs with droplet microresonators has been challenging primarily due to the difficulties in achieving stable coupling in liquids and precise mode manipulation. This severely hinders the design of flexible and integrated droplet laser devices for diverse applications such as biosensing. In addition, exploiting fully the characteristics of PMs requires dimensional tunability to precisely control the mode overlap even in real time, which is difficult for solid components commonly relying on complicated and rigorous fabrication. By comparison, benefiting from excellent flexibility and large-scale preparation capability, droplets are expected to be promising candidates as PM microlasers.

In this work, we report the first realization of such a liquid PM (LPM) microlaser based on coupled droplet microresonators and demonstrate its exceptional capability for ultrasensitive biosensing. Dye-doped oil droplets under pump of laser pulses are optically coupled in an aqueous environment via strong evanescent-field overlap (Fig. 1a), enabling the occurrence of super-modes extending both droplet microresonators by tailoring their spectral overlap. With strong mode suppression, typical single-mode lasing is observed in the LPM with a low threshold of ~610 nJ mm$^{-2}$. By incorporating photoisomerizable molecules into one of the coupled microresonators, dynamically tunable single-mode laser is achieved. We demonstrate that the spectral tuning exhibits the behavior of mode hopping, which features a magnified wavelength shift compared to that of individual droplets, enabling a nearly ten-fold spectral sensitivity enhancement. Remarkably, we propose a self-referenced intensity response mechanism that exploits the LPM lasing modes to detect small fluctuations induced by local refractive index changes at the cavity interface. Using this mechanism, we demonstrate ultrasensitive biomolecular sensing with a practical detection limit as low as 30 aM, three orders of magnitude lower than that of individual droplet microresonators, and a wide dynamic range over nine orders of magnitude. Our demonstration of the LPMs highlights unprecedented versatility and performance of the droplet microlasers, offering new prospects for droplet optofluidics and integrated biosensors.

## 2. Results

### 2.1 Design of LPM laser and biosensor

Our LPM lasers are designed to have a pair of droplets of slightly different sizes, as illustrated in Fig. 1a. The droplets are immersed in an aqueous environment and simultaneously pumped by a pulsed laser (532 nm). To realize active PMs, laser dye used as gain medium is incorporated into the droplets to provide effective optical gain. Upon irradiation by the laser pulse, WGM lasing is generated inside the droplet microresonators (hereafter droplets with larger and smaller diameters are termed, respectively, as µR$_1$ and µR$_2$). Adjacent droplets get coupled through spatially overlapped evanescent fields (Fig. 1b). For droplets of mismatched sizes, their resonance wavelengths (satisfying $2\pi n R_j = m_j \lambda_j$ with $j$=1, 2, where $m_j$ is an integer, $n$ and $\lambda_j$ are the effective refractive index (RI) and resonant wavelength of the mode, $R_j$ is the radius of the microresonator) are generally different, resulting in weak inter-resonator coupling. For this case the two droplets act as independent microlasers. Notably, by fine-tuning the droplet size, a specific lasing mode can be achieved at the wavelength where the microresonators are both on resonance (left panel in Fig. 1c). This suppresses all redundant lasing modes except the degenerate one, enabling single-mode laser emission from the coupled LPM. This is the so-called Vernier effect that has been widely studied in solid microresonators[27–31,33–36], but has yet to be demonstrated in droplets.

The merits of the mode selection are two-fold: it enables the LPM laser possessing a lower lasing threshold and significantly amplified spectral response to perturbations compared to common single droplets, both of which are highly desirable for biosensing applications. When environmental perturbations are imposed to a single droplet in the LPM (µR$_1$, for example), mode shifting occurs in its lasing spectrum due to the RI change (Fig. 1c, right panel). This spoils the original single-mode output while enabling mode hopping by tuning the relative resonance detuning and thereby shifting the Vernier overlap between both microresonators. Even a slight mode shift in µR$_1$ could lead to a significant mode hopping in the LPM. Therefore, in sharp contrast to the existing single droplet-based sensing schemes relying on multimode spectral shift[20–24], the proposed

LPM featuring single-mode hopping offers a significantly enhanced spectral sensitivity.

Next is how to convert the LPM lasers into efficient probes with sensing capability. As a proof of concept, a biosensing scheme for specific detection of biomolecular interaction is designed. The sensing mechanism is illustrated in Fig. 1d, in which intensity response of the LPM laser modes is monitored to identify biological binding events with high sensitivity. The single droplet with smaller diameter (i.e., $\mu R_2$) in the LPM is specifically modified to obtain various sensing functionalities. For example, specific antibody-antigen binding can be detected by functionalizing the droplet with well-designed biotin-streptavidin interactions (Fig. 1d, top left). Bridging by this specific molecular conjugation, resonance shift induced by slight antibody-antigen binding at the cavity interface within strong evanescent fields, albeit indistinguishable in the lasing spectrum, would degrade the Vernier overlap supporting the original single-mode emission (SM-1, Fig. 1d) while activating an adjacent mode (SM-2). Consequently, even a small RI variation induced by the target biomolecules can be greatly amplified. The resulting alternation in the lasing intensities of the LPM laser modes could serve as an ultrasensitive sensing signal output. Compared with conventional spectral-shift characterization, our strategy enables a much lower detection limit via monitoring the changes in the laser intensity signals. Moreover, our method establishes a unique self-reference mechanism through comparison of emission intensities among a series of laser modes. Therefore, common-mode noise sources (e.g., pump power drift and bulk RI changes) affecting all mode intensities in a similar, correlated manner could be cancelled. This goes beyond existing schemes that rely on monitoring average or spectrally integrated laser intensity in terms of accuracy and reliability.

**2.2 Low-threshold single-mode laser**

The lasing characteristics of the LPM were experimentally characterized under a far-field photoluminescence (PL) system (see Methods). A microfluidic chip was first employed to produce monodispersed dye-doped oil droplets in water with high uniformity and precisely determined sizes (Supplementary Note 1). Two Individual droplets of different diameters were extracted from the dispersion and then transferred to the same aqueous solution via a microcapillary (see Fig. 2a and Methods). Following interfacial contact, these droplets form stable adhesion at their junction, resulting in an LPM. Upon strong ns-pulse pumping, the coupled LPM exhibits typical single-mode laser emission (Fig. 2b). This indicates that strong coupling between the overlapping lasing mode can dramatically suppress all non-overlapping modes. Physically, non-overlapping modes are weakly coupled to the other microresonator, and can thus be considered as an additional loss mechanism. Since the lasing modes of the microresonators are highly size-dependent, wavelength modulation of the LPM lasers can be expected by simply varying the size ratio of droplets. As shown in Fig. 2b, single lasing modes are selected to lase depending on the geometrical parameters of the LPMs, resulting in flexibly tunable single-mode output within a range over 10 nm. In addition, a side-mode-suppression ratio (SMSR) exceeding 10 dB is achieved for the LPM lasers. These features are beneficial for precise optical sensing that causes peak wavelength shift.

Non-overlapping lasing modes are uncoupled and thus localized within individual microresonators. This contrasts sharply with the overlapping modes, where mode localization is significantly changed due to photon exchange. The optical fields are no longer localized instead changing into delocalized super-modes extending over both microresonators. This can be clearly observed in the dark-field PL graph of the LPM (inset in Fig. 2a), showing bright spots with nearly equal intensities at both opposing ends as well as at the central region of the LPM. These bright spots correspond intrinsically to the locations where the intracavity resonant mode field leaks most strongly into free space via radiation and scattering, i.e., out-coupling. When the resonance modes are spectrally matched, the delocalized super-mode exhibits pronounced field maxima at three characteristic locations (where out-coupling is strongest) along the LPM axis (Supplementary Fig. S2), thereby producing a characteristic emission pattern featuring dominant out-coupling hot spots. Once the modes become mismatched, they re-localize in the individual droplets and the number of participating modes increases, resulting in the emission pattern to degrade to multiple spots accordingly (see Supplementary Note 2 for more details). Moreover, the delocalized super-mode was found to have a lower lasing threshold compared to the laser modes in single droplets. Figure 2c compares the lasing thresholds for the single-droplet and the LPM lasers. The single droplet (characterizing multimode lasing) shows an explicit lasing threshold of ~6.5 µJ mm$^{-2}$, which is nearly ten-fold higher than that of the LPM, namely ~610 nJ mm$^{-2}$. Detailed statistical comparison of lasing thresholds is shown in Supplementary Note 3. The threshold reduction is primarily attributed to the formation of delocalized super-mode in the LPM that greatly decreases the multimode competition in the uncoupled droplets[4], thereby increasing the effective gain available to the lasing mode. This can be quantitatively analyzed through modelling the reduction in the number of competing modes (see Supplementary Note 4 for details). Notably, eigenmode simulations revealed that the coupled droplet modes exhibit a slightly decreased quality (*Q*) factor relative to the single-droplet mode (Supplementary Note 4), indicating the dominant role of the mode-selection effects responsible for the threshold reduction. Moreover, the increase in the effective gain length due to the coupling of resonators could also contribute to a lower threshold but to a lesser extent[46]. It should be noted that such a low threshold below 1 µJ/mm$^2$ remains unreported thus far for single-mode droplet lasers[37,38], which is also challenging for most multimode ones[21,39–45]. A comparison of the lasing thresholds for typical droplet microlasers is shown in Fig. 2d and detailedly presented in Supplementary Table S1.

**2.3 Tunable LPM laser and sensitivity enhancement**

To dynamically reconfigure the LPMs for tunable single-mode lasers, an *in situ* molecular isomerization strategy was introduced to tailor the effective RI of the microresonator. Experimentally, photoisomerizable spiropyran molecules were doped into the dye-doped oil solution (see Methods), followed by the generation of monodispersed spiropyran-doped oil droplets in preparation for LPMs. Only the larger droplet (i.e., µR$_1$) in the LPM was doped with the spiropyran molecules (Fig. 3a), for the purpose of triggering mode hopping. In contrast, dual modification of both droplets results in the unique behavior of single-to-multi mode switching (Supplementary Note 5). Under ultraviolet (UV) or visible (Vis) light, the spiropyran molecules isomerize from a closed-ring to a polar open-ring structure (or in turn)[30,31], which changes the RI of the oil solution (Fig. 3b). Upon

UV exposure of increasing time, continuous blueshifts of the lasing modes in the single spiropyran-doped droplet were experimentally observed (Fig. 3d), while those of another microresonator remained unchanged. This breaks the original mode matching and thus reconfigures the Vernier overlap (Supplementary Fig. S5), leading to continuous hopping of the single-mode lasing as illustrated in Fig. 3c. Moreover, the lasing tuning is fully reversible upon exposure to Vis light (Supplementary Fig. S6).

The LPM featuring single-mode hopping exhibits a magnified sensitivity versus the common single droplets. We quantify this by introducing a magnification factor (*M*-factor), which compares the wavelength shift of the LPM relative to that of the single droplet under identical UV exposure. The theoretical *M*-factor is given by $M=FSR_2/(FSR_2–FSR_1)$, where $FSR_j$ denotes the free spectral range (FSR) of $\mu R_j$ ($j$=1,2). For the LPMs consisting of 42-μm and 32-μm-diameter droplets, a mean *M*-factor of 4.9 during UV irradiation was experimentally achieved, as illustrated in Fig. 3e (left panel), which is close to the theoretical *M* of 4.8. Notably, the *M*-factor is closely related to the detuning of resonant wavelength between two microresonators, which trends to infinity as the detuning approaches zero. Therefore, it would be feasible to further enhance the *M*-factor by working with two droplets that are very close in size. As expected, the mean *M*-factor is increased when further reducing the size ratio, as shown in Fig. 3e. A maximum mean *M*-factor of 9.7 was experimentally obtained. This indicates that even a slight perturbation (caused by, e.g., bioreactions) inside a single droplet of the LPM could induce a significant amplification of the spectral wavelength shift with a clearly detectable mode hopping.

**2.4 Biosensing with LPM lasers**

The excellent tunable performance qualifies the LPM lasers as efficient biosensors capable of detecting various biological binding events at the droplet interface. As a proof-of-concept demonstration, we used the LPM lasers for detecting specific antibody-antigen binding kinetics, which is essential for immunoassay. To implement the intensity response mechanism for sensitive biomolecular detection (Fig. 1d), the smaller-diameter droplet in the LPM needs to be modified before being coupled to another (larger-sized) unmodified one. In the experiments, such a droplet was functionalized by employing the streptavidin-biotin interactions (see Methods and Supplementary Note 6 for details), considering its excellent affinity and specificity. The functional biotin groups enable the coating of the droplet with streptavidin conjugates that link biotinylated antibodies (anti-BSA) to the droplet interface. These specific binding processes occurring within strong evanescent fields of the droplet resonator can be identified from the lasing peak shifts (Supplementary Fig. S7). Control experiment was conducted to confirm these specific biomolecule interactions (Supplementary Note 7).

Then, a larger droplet without biomodification was prepared and coupled to the smaller, functionalized one. The observation of the single-mode laser emission validated the success construction of the LPM laser (Fig. 4a, right panel). The detection of antigen was subsequently performed based upon the functionalized LPM. We recorded the lasing spectra of the LPM laser when gradually increasing the BSA concentration from aM to nM, as presented in the right panel of Fig. 4a. With the successive injection of the analyte solution with incremental biomolecular concentrations, the lasing spectra exhibited the representative mode hopping. This remarkable response in the laser signal fully demonstrated specific antibody-antigen binding at the cavity interface (Fig. 4a, left panel). Importantly, distinctive intensity changes of the LPM laser peaks occurred during the incubation (interaction) process, enabling an ultralow limit of detection (LOD).

Figure 4b shows the spectral response of the LPM at various analyte concentrations, indicating an intensity alternation of the three lasing modes (SM-1, -2, and -3) with increasing the amounts of analyte. The intensity response was more sensitive to the analytes at lower concentrations; For higher analyte concentrations, this sensitivity exhibited a distinct slowing trend, probably due to the reduction in binding sites of the capture antibody. A clear reduction in the intensity ratio of the two adjacent lasing modes (e.g., the original line SM-1 to the hopped laser line SM-2) was observed after the initial analyte injection. The measurement indicates a reliable detection of biomolecules down to 30 aM, which can be determined using the three-sigma method (see Methods for details). By comparison, the limit of BSA detection for the single droplet laser was tested to be ~30 fM (Supplementary Note 8). This reveals that our LPM biosensor accomplished a three-order-of-magnitude improvement in the LOD, highlighting the potential of the LPM platform for high-performance biosensing. We mention that the improvement of LOD is primarily due to the switch from measuring a small, absolute wavelength shift to monitoring a large, relative intensity change between two laser modes. The intensity response is inherently more sensitive than measuring the absolute position of a broadened laser peak in a single droplet. Therefore, the large $M$-factor ensures that a tiny perturbation triggers a complete and easily detectable mode hopping, whereas the intensity ratio ensures the read out of a tiny perturbation with high resolution and precision. Moreover, the intensity response measurement also exhibited excellent robustness against common-mode noise sources, manifesting as stable intensity ratio versus, e.g., the pump power fluctuations (see more details in Supplementary Note 9).

Our biosensor can operate in a wide dynamic range over nine orders of magnitude by monitoring the intensity response of the LPM laser modes. For various analyte concentrations ranging from aM to nM, the mode intensities were extracted from the LPM laser spectra and plotted versus the analyte concentration in color scale (see right inset in Fig. 4b). With these series of lasing modes simultaneously monitoring the specific biomolecular binding kinetics, we can readily recognize arbitrary analyte concentrations in a wide dynamic range but with a higher resolution compared to the spectral shift detection. By examining various combinations of antibodies and antigens (see Methods), our sensor exhibited a high specificity to the target analyte (see Fig. 4c and Supplementary Figs. S11 and S12). In addition, theoretical fitting of the spectral response curve validates the specific molecular binding kinetics characterized by the observed spectral shifts (Fig. 4d, see Methods and Supplementary Note 10 for details of theoretical analysis).

## 3. Discussion

In brief, we demonstrated the realization of the LPM microlasers with optically coupled droplet microresonators that enable low-threshold lasing and ultrasensitive biomolecular sensing. Beyond the present results, our demonstration of the LPM establishes a versatile photonic platform with broad applicability and universality (see Fig. 5). With exceptional lasing characteristics and sensing capabilities, the LPM microlaser is expected to be a promising candidate as microprobes for biosensing applications both *in vivo* and *in vitro*.

Particularly, using micro-injection techniques to implant biocompatible droplets into live cells and organisms for cellular monitoring and tracking has received intense interest recently[21,22,24]. Our approach opens up new avenues forward for further equipping these techniques with higher detection sensitivity and better laser emission performance, promising to overcome the difficulties of strong background signals and high scattering and absorption in complex biological media. Moreover, the design principles of our LPM microsensor could inherently address the key challenges in such complex environments. Its self-referenced intensity-ratio mechanism offers inherent resilience to bulk RI fluctuations common in biological fluids. The surface functionalization strategy could effectively mitigate non-specific adsorption of proteins or molecules on the droplet surface. Our technique could also be extended to medical diagnosis applications, where LPMs could not only serve as implantable sensors for detecting health biomarkers, but also be integrated into lab-on-chip devices to analyze biological samples with ultrasmall detection volumes. A more detailed discussion regarding microfluidic integration for high-throughput assays can be seen in Supplementary Note 11.

In addition, building on the rapid advances in droplet bioreactors[19], LPMs that encapsulate various bioactive materials (e.g., biomolecules, bacteria, and single cells) enable direct interactions between bioreactions and the primary resonating energy, potentially providing more sensitive laser signal responses to bioactivity analysis inside droplets. Beyond these bioapplications, we envision that additional insight could be gained by extending our framework to broader physical realms, such as non-Hermitian physics, cavity optomechanics[47,48], and metasurfaces[49]. This would open up exciting prospects to pioneering unexplored domains of non-Hermitian droplet photonics, droplet optomechanical devices, and hybrid droplet-metasurface systems.

## Methods

### Droplet preparation

A customized microfluidic droplet generator chip was used to generate highly monodispersed dye-doped oil droplets in DI water. The continuous phase during production of the droplet was an aqueous surfactant solution. Sodium dodecyl sulfate (SDS, Aladdin) was used as the surfactant (except for the biosensing experiments) and was added to the water phase at a concentration of 6 mM to stabilize and store the droplets. The disperse phase is a mixture of oil and dye. Edible sunflower oil (RI=1.474, Aladdin) was selected as the droplet material considering its good biocompatibility. Organic dye Bodipy-2 (Aladdin) was dissolved in the oil at a concentration of 1 wt%. The flow in each channel of the microfluidic chip was controlled using a commercial syringe pump. By changing the pressures of the water and oil phases, the droplet size was precisely controlled in the size range of 30-50 μm (Supplementary Fig. S1). Different sizes of droplets were collected in separate vials to be later used for LPMs. For the fabrication of all-optical tunable droplet lasers, photoisomerizable molecules 1',3'-Dihydro-8-methoxy-1',3',3'-trimethyl-6-nitrospiro[2H-1-benzopyran-2,2'-[2H]indole] (Aladdin) were specifically doped into the oil/dye solution at a concentration of 1 wt%, which was absent in the biosensing experiments.

A spectral pre-screening procedure was performed to prepare the optically coupled droplets (i.e., LPMs). The single oil droplets with different diameters were separately extracted with a microcapillary connected to a syringe (Fig. 2a). The microcapillary was then discharged into the aqueous surfactant solution placed in a petri dish. Each droplet (of different diameters) was tested individually under identical pumping conditions, and its WGM lasing spectrum was accordingly recorded. The droplets of two target sizes satisfying the Vernier overlap were selected by checking their lasing spectrum. For example, two droplets with diameters of ~32 μm and ~42 μm were experimentally identified for generating the coupled LPM, based on the fact that their lasing spectra shared a common lasing mode (Supplementary Fig. S16). The spectrally pre-screened droplets were finally brought into gentle contact using an optical fiber tip that may induce minor deformations at the moment of contact but did not compromise the optical performance and therefore affect the subsequent measurement. Optical coupling of the droplets was identified once a clean single-mode lasing peak was observed. Statistic measurement on achieving single-mode lasing across multiple, independently fabricated droplet pairs demonstrated high reproducibility (~68%, see Supplementary Fig. S17) of the formation of stable LPMs. The coupled LPMs could also be formed using an integrated microfluidic chip involving an inverted microwell array (see Supplementary Note 11 for details). During all sensing experiments, the petri dish containing oil droplet solution was covered by a transparent glass lid to minimize the impact of water evaporation on the bulk RI fluctuations in the surrounding medium of the oil droplets.

**Experimental setup**

Optical characterization of the droplet lasers was performed under a far-field PL system (see schematic illustration in Supplementary Fig. S16a). The system was equipped with a pulsed laser (Beamtech Optronics, Nimma-900) integrated with an optical parametric oscillator, an inverted optical microscope (Nikon, Ti2), a CCD camera (Nikon, DS-Fi3), and a spectrometer (Zolix, Omni-λ300i) equipped with an EMCCD (Andor, DU-897U). The pulsed laser has a repetition rate of 10 Hz and a pulse duration of 7 ns, and its pump wavelength was set to 532 nm to efficiently excite the Bodipy-2 dye. The pump pulses were focused onto the sample under an incident angle of 45° with respect to the substrate normal using a lens. The pump spot was chosen to have a diameter of ~2 mm on the sample surface to homogeneously pump the LPM. The PL emission was collected perpendicular to the LPM with a microscope objective (20×, NA=0.45) and a long-pass filter, which was then either imaged on the CCD camera or analyzed in the spectrometer (grating with 1200 lines $mm^{-1}$) equipped with the EMCCD with an overall spectral resolution of ~0.17 nm. The RI of the photoisomerizable molecule-doped oil was measured on an Abbe refractometer with a resolution of $1\times10^{-4}$ RIU. A mercury lamp was used to deliver UV and Vis light to induce photoisomerization in the droplets.

**Droplet functionalization for binding of biomolecular analytes**

For constructing the LPM biosensors, we prepared two sets of oil droplets with similar diameters of approximately 32 and 35 μm using the microfluidic chip. The droplets were generated in aqueous surfactant solution to control their interfacial tension. Non-ionic and non-denaturing Pluronic F68 (Sigma-Aldrich) block polymer surfactant was chosen to replace SDS since the latter could denature streptavidin and abolish its biotin-binding activity. The 32-µm-diameter droplets were pre-functionalized with streptavidin-biotin interactions, while the larger-sized ones kept unmodified. Specifically, we functionalized the droplets in three steps (see Supplementary Note 6). Firstly, the lipid-containing oil was obtained by dilution of DSPE-PEG-Biotin (Aladdin) phospholipids in sunflower oil at 330 µM, followed by 30 min of sonication and evaporation of the ethanol from the oil at room temperature. The biotinylated oil droplets were then generated in an aqueous continuous phase containing 40 µM Pluronic F68 surfactant at critical micelle concentration (CMC). Afterwards, a single droplet was extracted from the dispersion and delivered into a lidded petri dish containing 5 mL aqueous solution of Pluronic F68 at CMC using a microcapillary. After the first step of insertion of biotinylated lipids at the droplet interface, 0.2 mL streptavidin (Aladdin) aqueous solution at 75 µM was added via a micropipette, followed by an incubation time of 20 min and a washing (diluting) step of 5 min with PBS to remove excess streptavidin. During the incubation, streptavidin attachment onto the biotinylated droplet surface was monitored by tracing the lasing wavelengths of the droplet laser (see Supplementary Fig. S7). In addition, the washing does not affect the droplet spectra as confirmed by stable lasing wavelengths during the washing process. Finally, 20 µL biotinylated rabbit anti-BSA (purchased from OKA) with a concentration of 1mg mL$^{-1}$ was added into the droplet solution, followed by an incubation time of 30 min during which successive measurements of droplet lasing spectra were performed at 5 min intervals. The residual anti-BSA molecules were removed with PBS washing for 5 min.

Next, the larger droplet (diameter 35 µm) without biomodification was prepared and delivered into the same solution to couple it to the smaller one. To ensure long-term stability of droplet coupling, 1% w/w sodium alginate (Aladdin), widely used as a biocompatible and relatively low-fouling polysaccharide in bioassays, was added to the bulk solution without interfering with biosensing. A brief 10-min incubation sufficed to maintain stable droplet coupling for over 12 hours, enabling continuous measurement of single-mode lasing outputs throughout this period of time (see Supplementary Note 12 for stability test). For the detection of antigens, the BSA (OKA) solution at various concentrations were prepared and added into the droplet solution in steps. For certain amounts of analyte, stable intensity outputs of the LPM laser can be achieved once the dynamic equilibrium was established upon the balanced binding and dissociation processes. Therefore, the addition of each concentration was determined based on the stabilization of spectral intensity at least for 30 min.

For specificity testing experiments, various combinations of the antibodies rabbit anti-BSA, rabbit anti-goat IgG, goat anti-human IgG, and the antigens BSA, goat IgG, human IgG (all purchased from OKA) were examined. The functionalization of the antibodies was the same as the operation mentioned above, and the antigens were prepared with the same concentration. Each antigen was injected into the droplet solution reaching a final concentration 30 nM, followed by incubation for 1 hour during which the lasing wavelength shift was measured. The total wavelength shift was recorded after 30 min of spectral stabilization.

**Determination of LOD**

The LOD was determined using the three-sigma criterion based on blank measurements[50,51]. To be specific, low levels of BSA ranging from 0 to 30 aM were injected while the intensity ratio, $I_{SM-2}/I_{SM-1}$, was measured from three independent experiments (Supplementary Fig. S19). For the BSA concentration of 0 aM, the mean ±3 standard deviation (SD) values of $I_{SM-2}/I_{SM-1}$ were obtained from repeated buffer (i.e., blank) injections, giving a detection threshold of

$$R_{th} = \mu_{blank} + 3\sigma_{blank} \tag{1}$$

where $\mu_{blank}$ and $\sigma_{blank}$ are the mean and SD of the blank response. For other analyte concentrations, the mean response $R(C)$ was measured and then compared with $R_{th}$. A concentration was considered detectable when $R(C) > R_{th}$; that is, the experimental LOD was defined as the lowest tested concentration satisfying this condition. Upon substituting the blank statistics into Eq. (1), the 3σ threshold was calculated to be $R_{th} \approx 0.0192$, giving an approximate LOD of 30 aM, as illustrated in Supplementary Fig. S19.

**Theoretical model for biomolecular detection**

To elaborate on the origin of the observed spectral response and its evolution with the analyte concentration, we established a theoretical model that describes the wavelength shift induced by the specific antibody-antigen binding (see Supplementary Note 10 for more details). Theoretically, analytes adsorbed on the cavity surface introduce a wavelength shift $\Delta\lambda$, which could be obtained by

$$\Delta\lambda = \frac{\lambda}{n_{eff}} \frac{dn}{dc} \frac{\Gamma}{\delta} f\eta g \tag{2}$$

where $dn/dc$ is the RI increment of the biomolecules, $\Gamma$ the surface mass density, $f$ the polarization amplification factor that characterizes the sensitivity of interface RI disturbance to different polarization modes[52], $g$ the evanescent-field axial weighting factor that is introduced to correct the attenuation effect caused by molecular lifting[53], $\delta$ and $\eta$ the evanescent penetration depth and the evanescent volume fraction that can be estimated by numerical analysis (see Supplementary Fig. S13). Details on the calculations of these quantities and the associated derivation of wavelength shift are presented in Supplementary Information. Based on Eq. (2), the peak shifts of the droplet laser were fitted to the Hill-Langmuir specific binding model[54], yielding a binding curve as illustrated in Fig. 4d.

**Numerical simulation**

We carried out 2D numerical simulations using finite element method via COMSOL Multiphysics to model the electric field distribution of the coupled droplet resonators. The simulation parameters are detailed as follows. The refractive indices of the droplet and the surrounding medium were set to 1.474 (oil) and 1.33 (water), the droplet diameters were 42 μm and 32 μm, the gap between two droplets was 100 nm considering the presence of surfactant molecules, the excitation wavelength was near 585 nm, and a perfectly matched layer and scattering boundary conditions were used as the external boundary of the model.

## Data availability

Source data are provided with this paper, and more detailed data are available from the corresponding author upon request.

## Code availability

The codes used in this paper are available from the corresponding author upon request.

## Acknowledgements

This work was supported by the National Key R&D Program of China (Grant No. 2024YFE0102400 H.J.); The National Natural Science Foundation of China (NSFC) (Grant Nos. 12421005 H.J., 12247105 L.-M.K., 12175060 L.-M.K., 12205256 Y.W., 62571494 J.-L.W., 12304407 J.-L.W., 62505201 R.D., 12405029 Y.-F.J.); CPG2025-00034 H.-D.S., SRG2023-00025 H.-D.S., and the Science and Technology Development Fund from Macao SAR (FDCT) (0002/2024/TFP H.-D.S.); The Hunan Major Sci-Tech Program (Grant No. 2023ZJ1010 H.J.); The Innovation Program for Quantum Science and Technology (Grant No. 2024ZD0301000 H.J.); The Science and Technology Major Project of Henan Province (Grant Nos. 241100210400 H.J., 231100220800 L.-Y.J.); Joint Fund of Henan Province Science and Technology R&D Program (Grant No. 225200810071 L.-Y.J.); The Key Research and Development Projects of Henan Province (Grant No. 241111222900 L.-Y.J.); HNQSTIT project (Grant No. 2022112 H.J.); The Natural Science Foundation of Henan Province (Grant No. 252300421221 Y.-F.J.). The authors thank Prof. Lei Shi for his insightful suggestions on theoretical analysis.

**Author contributions**

Y.W. and Y.-H.H. designed and performed the experiments. Y.W., Y.-H.H. and J.-L.W. provided theoretical analyses. Y.W. and Y.-H.H. designed the figures and wrote the manuscript with contributions from all authors. R.D., Y.-F.J., H.-Y.W., L.-Y.J., L.-M.K. and H.-D.S. participated in discussions. H.J. guided the research and supervised the project.

**Competing interests**

The authors declare no competing interests.

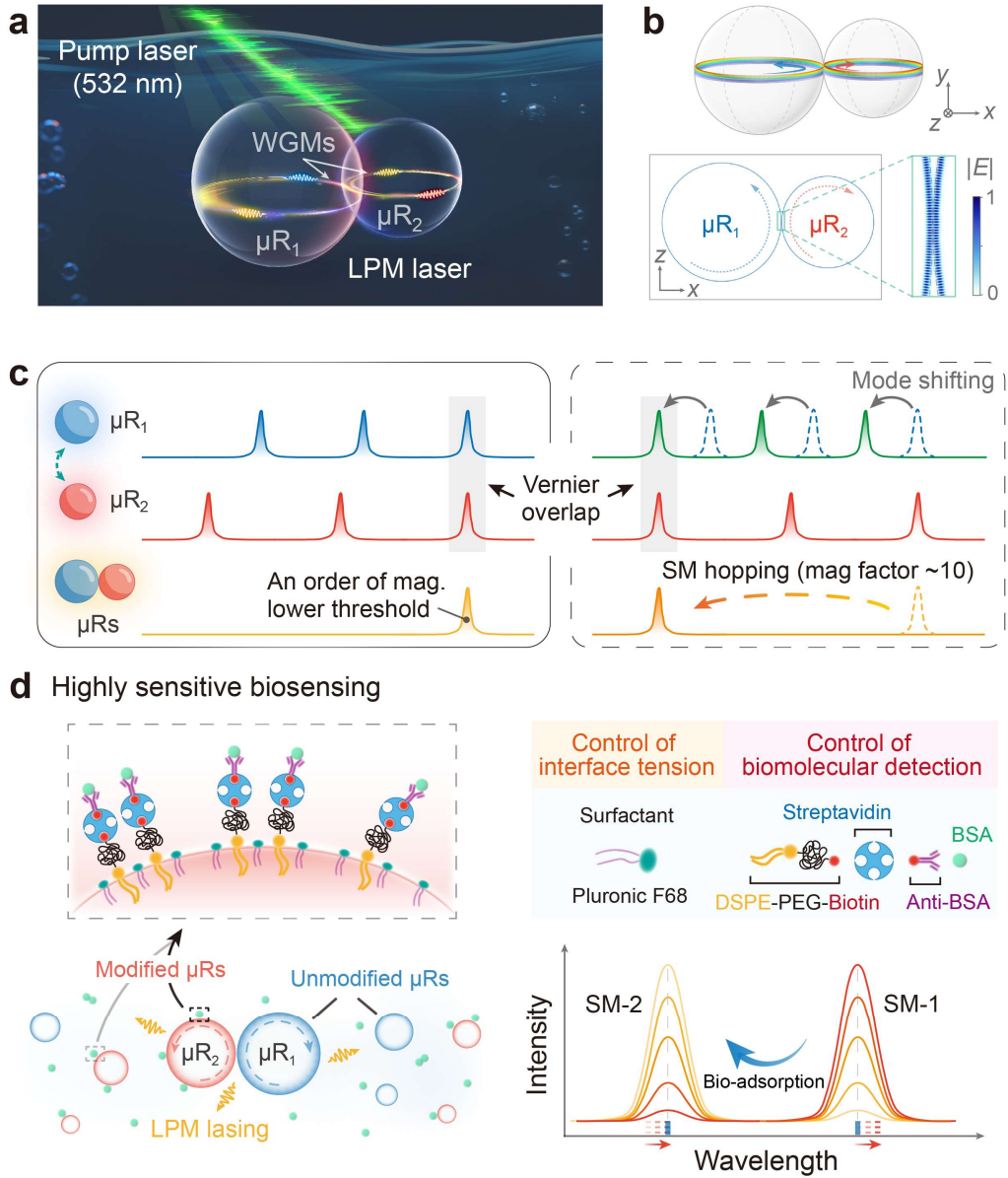

**Fig. 1. Configuration of LPM laser for biosensing. a** Schematic of an LPM laser consisting of two size-mismatched dye-doped droplets immersed in an aqueous environment. A 532 nm pulsed laser is used to excited the optical WGMs of the droplet microresonators (μRs). Adjacent droplets are optically coupled through spatially overlapped evanescent fields. **b** Numerical simulation of the electric field distribution in the coupled droplet μRs with diameters of 42 (μR$_1$) and 32 μm (μR$_2$). The amplitude of the electric field |E| distributed on the *x-z* plane is plotted for the coupled WGMs, where strong evanescent field overlaps and photon exchange occurs at the gap region. Detailed simulation parameters are given in Methods. **c** Illustration of the LPM supporting a lower (an order of magnitude (mag.)) threshold single-mode (SM) laser emission (left panel) and an amplified mode shift (right panel) compared to single droplets. Reconfiguration of the Vernier overlap between specific lasing modes of both droplet μRs enables a significant mode hopping with a magnification (mag) factor of ~10. **d** Schematic concept of the LPM laser as a highly sensitive biosensor. Intensity response of the LPM laser modes serves as sensitive sensing signal output, which is utilized to identify biological binding events at the cavity interface.

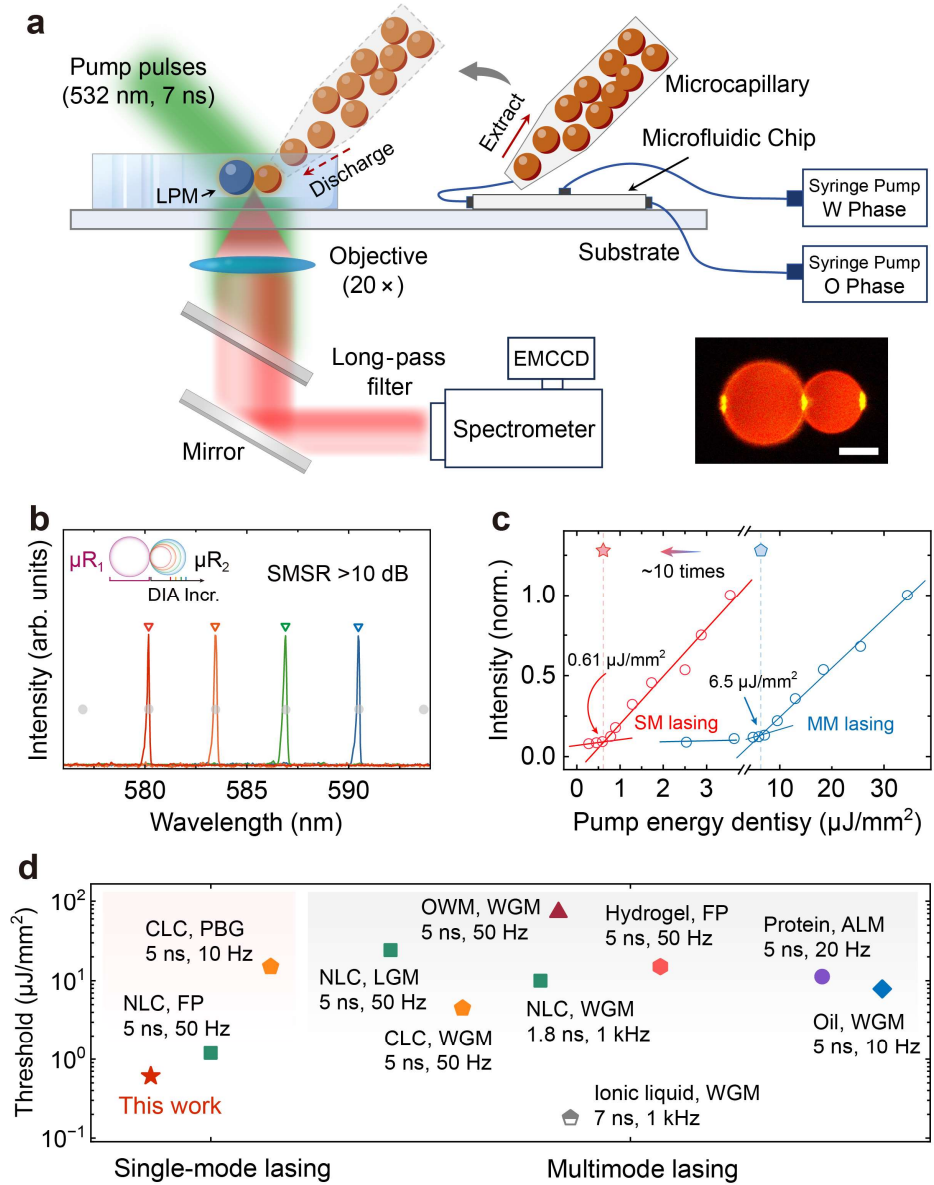

**Fig. 2. Optical characterization of the droplet microlasers. a** Schematic diagram of the experimental setup for droplet preparation and spectral characterization. Dimensions are not to scale. Monodispersed dye-doped oil droplets are produced using a microfluidic chip and transferred via a microcapillary. Droplets dispersed in the water are irradiated by a pulsed laser (532 nm, 7 ns) to excite the WGM lasing. PL spectrum is collected via an objective and then detected by a spectrometer equipped with an EMCCD. The right inset shows the dark-field PL graph of the LPM with diameters of ~42 and ~32 μm. Scale bar, 20 μm. **b** Single-mode lasing spectra of the LPMs with varied diameter ratios schematically shown in the left inset, showing a side-mode-suppression ratio (SMSR) exceeding 10 dB. The diameter of μR$_1$ is fixed to ~42 μm, while the diameters of μR$_2$ are ~29.5, ~32, ~35.4, and ~38.5 μm, respectively. "DIA Incr." denotes diameter increment of μR$_2$. The gray circles mark the lasing peaks of the 42-μm-diameter single droplet. **c** Threshold curves of both a single droplet exhibiting multimode (MM) lasing and an LPM exhibiting single-mode (SM) lasing. **d** Comparison of single-mode and multimode lasing thresholds for representative droplet microlasers reported in refs. [21,37–45]. Material, mechanism, and main pump-laser parameters (pulse duration and repetition rate) of the droplet microlasers are marked in the plot. NLC, nematic liquid crystals; CLC, cholesteric liquid crystals; OWM, oil-water mixture; F-P, Fabry-Perot; PBG, photonic band gap; LGM, Laguerre-Gaussian modes; ALM, Arc-like modes.

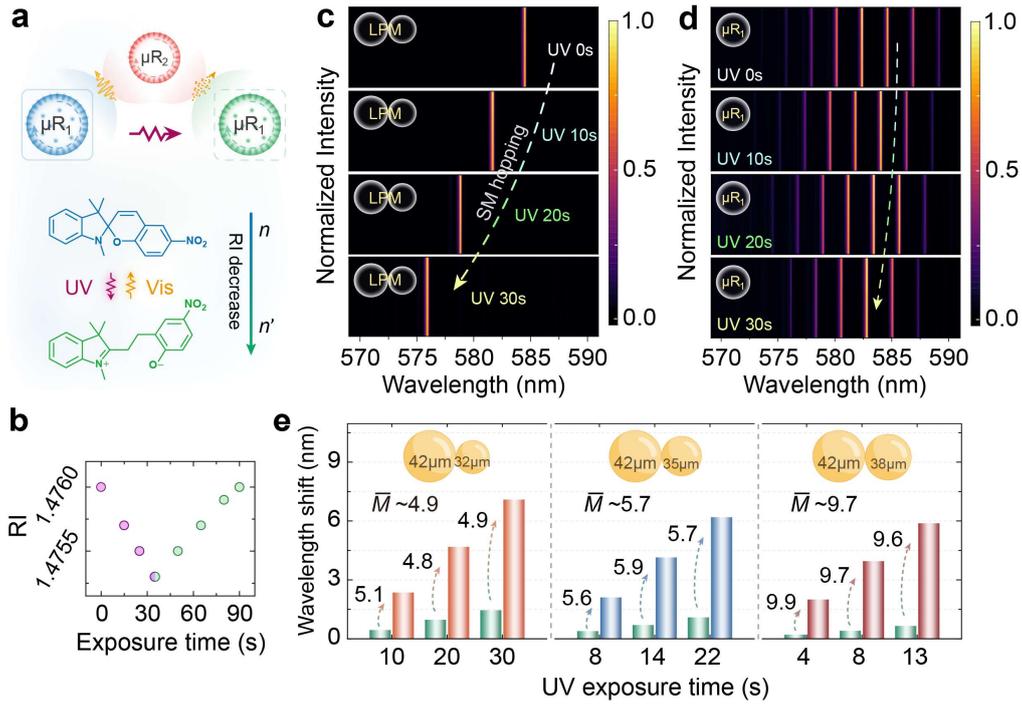

**Fig. 3. Optically controlled single-mode lasing and sensitivity enhancement. a** Illustration of the molecular isomerization strategy for constructing tunable LPM laser. The bottom panel shows reversible photoisomerization of the spiropyran molecule by irradiation with ultraviolet (UV) and visible (Vis) light. UV irradiation reduces the compatibility between the ring-opened spiropyran and the nonpolar oil solution, leading to localized molecular aggregation. The formation of the aggregated states decreases molecular free volume and overall medium polarizability, resulting in overall decrease of the RI of the oil solution. **b** Measured refractive index (RI) of the spiropyran-doped oil solution (volume 20 μL) versus the exposure time to UV and Vis light. **c** Mode hopping of the LPM consisting of two coupled droplets with diameters of 42 (μ$R_1$) and 32 μm (μ$R_2$), under UV exposure of increasing time. **d** Lasing spectra of the single droplet (μ$R_1$) under successive excitation of UV light. **e** Comparison of wavelength shifts between the single droplet (μ$R_1$) and different LPMs with varied size ratios as schematically illustrated in the insets. The corresponding *M*-factors characterizing sensitivity enhancement are labelled in the plots. The lasing wavelengths in (**e**) are extracted by fitting the spectral peaks to Lorentzian curves.

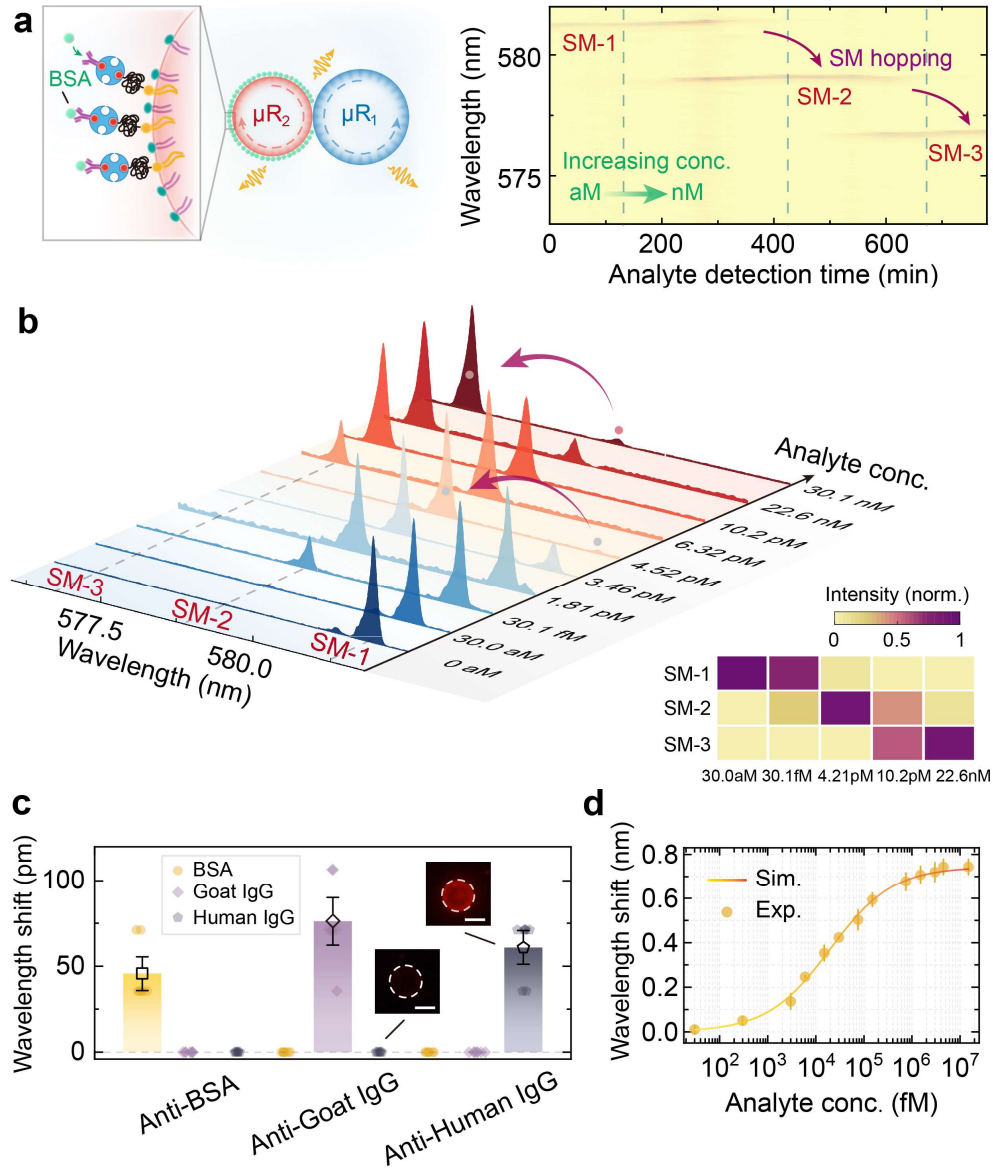

**Fig. 4. Demonstration of biosensing by LPM lasers. a** Schematic illustration of specific biomolecule binding at the droplet interface (left panel) and peak wavelength dynamics of the LPM laser (right panel) during biomolecular detection process. SM hopping occurs twice during increasing the analyte concentration, indicating dynamic reconfiguration of the Vernier overlap. **b** Intensity response of the LPM laser modes with increasing the analyte concentrations (conc.). SM hopping occurs at concentrations of 4.52 pM (SM-1 to SM-2) and 30.1 nM (SM-2 to SM-3). The right inset shows two-dimensional color map of the laser mode intensities under different analyte concentrations. All mode intensities at each concentration are normalized by the intensity sum of the three mode intensities. **c** Wavelength shifts of the single antibody-functionalized droplets after incubation with different antigens, showing excellent specificity to the target analyte. Error bars are standard deviation (SD) from seven independent experiments. The insets show the fluorescence images of the droplets (see more details in Supplementary Fig. S11). Scale bar, 20 μm. **d** Spectral response curve for biomolecular binding. Shown are experimental (circles) and simulated (curve) results of the wavelength shift of the single antibody-functionalized droplet versus the antigen concentration. Error bars are SD from three independent experiments.

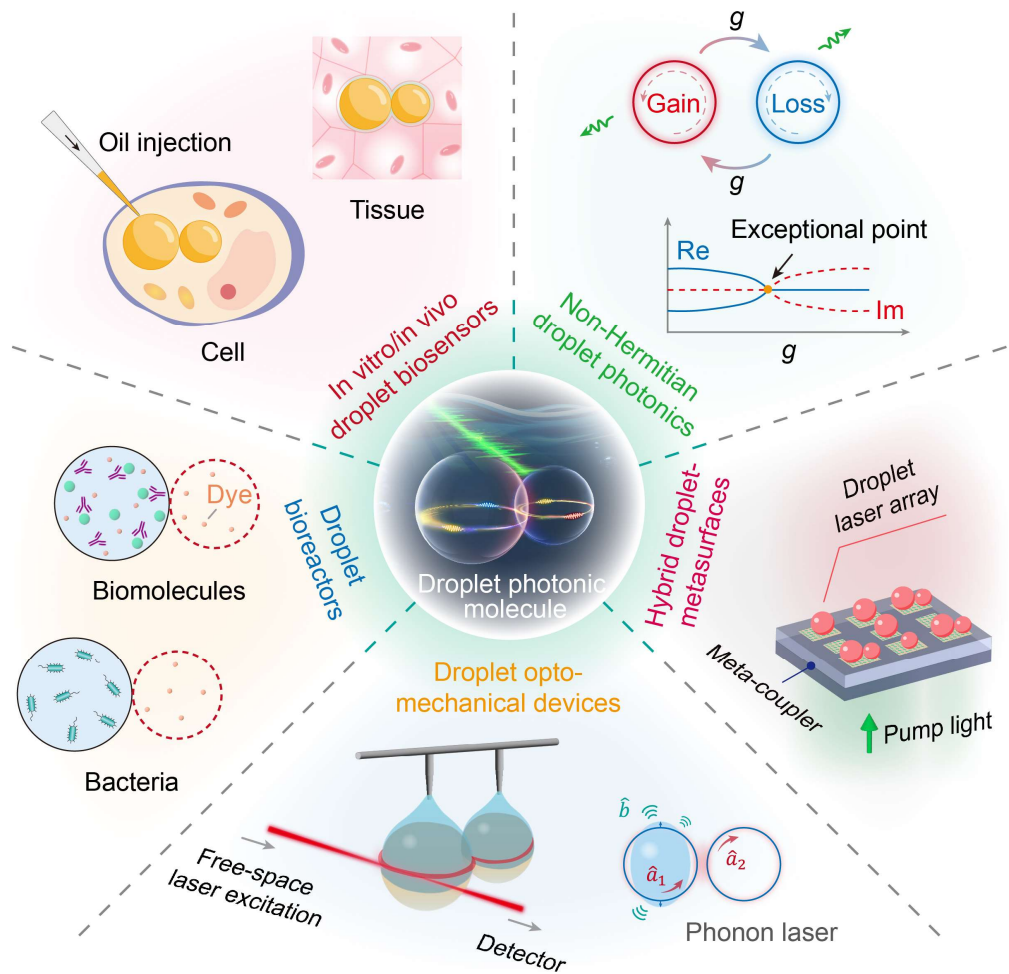

**Fig. 5. Broad applicability and universality of the LPM.** The proposed photonic structure paves a general route to exploring a range of fundamental physical effects and applications, including non-Hermitian droplet photonics, hybrid droplet-metasurfaces, droplet optomechanical devices, droplet bioreactors, and in vitro or in vivo droplet biosensors.